\documentclass{article}
\usepackage{graphicx}
\usepackage{amsmath,amssymb,amsfonts}
\usepackage{xcolor}
\usepackage{hyperref}
\usepackage{textcomp}
\usepackage[capitalize]{cleveref}
\usepackage{subcaption}

\usepackage[maxbibnames=2,isbn=false,url=false,doi=false,eprint=false,giveninits=true,sorting=none,style=numeric-comp]{biblatex}
\usepackage{tikz}
\usetikzlibrary{3d,decorations.pathmorphing,shapes.symbols,shapes.arrows,through}
\tikzset{positron/.style={
			->,
			blue,
			decorate,
			decoration={snake,amplitude=.75,segment length=2mm}
		},
	axis/.style={
			->,
			black!50
		}}

\usetikzlibrary{calc, angles, quotes, arrows.meta, decorations.pathreplacing}

\definecolor{hitblue}{RGB}{55,138,221}
\definecolor{validgreen}{RGB}{29,158,117}
\definecolor{invalidred}{RGB}{216,90,48}
\definecolor{unitgray}{RGB}{130,130,130}

\usepackage{acro}
\acsetup{single}
\DeclareAcronym{2D}{short=2D,long=two-dimensional,first-style=short}
\DeclareAcronym{3D}{short=3D,long=three-dimensional,first-style=short}
\DeclareAcronym{CRT}{short=CRT,long=coincidence resolving time}
\DeclareAcronym{LFOV}{short=LFOV,long=large field-of-view}
\DeclareAcronym{MAP}{short=MAP,long=maximum a posteriori}
\DeclareAcronym{MLEM}{short=MLEM,long=Maximum-Likelihood Expectation-Maximization}
\DeclareAcronym{NM}{short=NM,long=Nelder-Mead}
\DeclareAcronym{PET}{short=PET,long=positron emission tomography}
\DeclareAcronym{Ps}{short=Ps,long=positronium}
\DeclareAcronym{o-Ps}{short=o-Ps,long=ortho-positronium}
\DeclareAcronym{p-Ps}{short=p-Ps,long=para-positronium}
\DeclareAcronym{PDF}{short=PDF,long=probability density function}
\DeclareAcronym{PLI}{short=PLI,long=positronium lifetime imaging}
\DeclareAcronym{QED}{short=QED,long=quantum electrodynamics}
\DeclareAcronym{TOF}{short=TOF,long=time-of-flight}
\DeclareAcronym{TRIO}{short=TRIO,long=Three-Photon Bayesian Imaging of Ortho-Positronium}

\title{Ortho-Positronium Three-Photon Decays: Physics Constraints and a Closed-Form Energy Method for Annihilation Vertex Reconstruction}
\author{L. Raczyński\thanks{L. Raczyński, M. Bała, K. Klimaszewski, M. Obara, R. Y. Shopa are with the National Centre for Nuclear Research, Department of
        Complex Systems, 05-400 Otwock, Poland (e-mail: lech.raczynski@ncbj.gov.pl).},
    W. Krzemień\thanks{W. Krzemień is with the National Centre for Nuclear Research, Department of
        High Energy Physics, 05-400 Otwock, Poland (e-mail: wojciech.krzemien@ncbj.gov.pl).},
    A. Coussat\thanks{A. Coussat is with INSA-Lyon, Université Claude Bernard Lyon 1, CNRS, Inserm, CREATIS UMR 5220, U1294, F-69373, Lyon, France (e-mail: aurelien.coussat@creatis.insa-lyon.fr).},\\
    M. Bała,
    B.C. Hiesmayr\thanks{B. C. Hiesmayr is with the IT:U Interdisciplinary Transformation University, Freistädter Strasse 400, 4040 Linz, and University of Vienna, Faculty of Physics, Währingerstrasse 17, 1090 Vienna, Austria.},\\
    K. Klimaszewski,
    M. Obara,
    R. Y. Shopa}
\date{April 2026}

\begin{document}

\maketitle

\begin{abstract}
We examine the physical foundations of  \acl{o-Ps} three-photon decay in the context of annihilation vertex reconstruction, focusing on how energy–momentum conservation constrains the space of physically admissible solutions. Finally, we provide a closed-form analytical derivation of an energy-based vertex reconstruction algorithm.
\end{abstract}

\section{Introduction}

Positronium is one of the simplest purely leptonic bound states known in nature, consisting of an electron and a positron (the antimatter partner of the electron) held together by their mutual Coulomb attraction. The solution of the Schrödinger equation describing this quasi-atom is the same as for the hydrogen atom, except that Bohr's radius is four times greater.
Positronium decays via annihilation of the constituent positron either with its bound electron or with electrons from the surrounding medium.
It exists in two spin configurations: \acf{p-Ps}, with a lifetime in vacuum of 124\,ps, which predominantly decays into two photons, and \acf{o-Ps}, which decays mainly into three photons and has a mean lifetime in vacuum of 142\,ns~\cite{oreThreePhotonAnnihilationElectronPositron1949}.
Positronium and its properties have been the subject of various fundamental studies, including precision tests of \ac{QED}~\cite{adkinsPrecisionSpectroscopyPositronium2022}, measurements of discrete symmetry violations~\cite{PhysRevLett.91.263401, PhysRevLett.104.083401,allenMeasurementPositroniumDecays2025}, investigations of correlations in photon polarization degrees of freedom ~\cite{hiesmayrGenuineMultipartiteEntanglement2017,moskalFeasibilityStudiesPolarization2018a, hiesmayrQuantumErrorChannels2024a,parashariClosingDoorPuzzle2024, caradonnaKinematicAnalysisMultiple2024a, caradonnaStokesparameterRepresentationCompton2024, bordesFirstDetailedStudy2024, balaProbingArbitraryPolarized2025, zugecReconciliationPryceWardKleinNishina2026,kozuljevicPolarizationenhancedPETStudy2026} and  positronium’s wave nature~\cite{nagataObservationPositroniumDiffraction2025}.

Since annihilation photons carry information about both the positronium system itself and the microscopic environment in which it annihilates, positronium serves as a sensitive probe of biological and material environments.
For instance, \ac{Ps} is used in material science and engineering studies, where the PALS technique provides direct information about defect structures in materials~\cite{jean2003principles}.
Recently, the positronium-based marker has been discussed and studied in the context of \ac{PET} medical imaging~\cite{hourlierExperimentalUsesPositronium2024}.
In the conventional \ac{PET} imaging, the functional image of the patient is obtained by reconstructing the spatial distribution of the radiotracer concentration from photon pairs produced by annihilation of positrons emitted by an administered radiotracer.
By measuring properties of the positronium system, which precedes annihilation in biological tissues in approximately 40\% of cases~\cite{harpenPositroniumReviewSymmetry2003}, additional diagnostically relevant information about the local environment can be extracted.

The proposed markers include: (i) the positronium lifetime~\cite{moskalPositroniumImageHuman2024a, shopaPositronium2023, steinbergerPositroniumLifetimeValidation2024a, huangFastHighresolutionLifetime2025,mercolliFirstPositroniumLifetime2025,mercolliPhantomImagingDemonstration2025,takyuPositroniumLifetimeMeasurement2024,huangHighResolutionPositroniumLifetime2025}, which is shortened by interactions with the surrounding medium through pick-off annihilation~\cite{brandtPositroniumDecayMolecular1960, trungInvestigationOrthopositroniumAnnihilation2023} and spin-exchange processes~\cite{zgardzinskaOrthoPara2015, stepanovInteractionPositroniumDissolved2020}, and (ii) the ratio of number of three to two photon decays.
The latter has been studied as a diagnostic observable sensitive to  microstructure~\cite{kacperskiThreegammaAnnihilationImaging2004}, and has also found use in materials science for the characterisation of porous media and gas diffusion~\cite{pevovarRatioPositronAnnihilation2007, kauppilaInvestigationsPositroniumFormation2004}.
A broader overview of positronium applications in biology and medicine is provided in~\cite{hourlierExperimentalUsesPositronium2024}.

Two main classes of three-photon vertex reconstruction methods have been proposed.
The first and earliest relies on energy-momentum conservation to derive the vertex position from the measured photon energies~\cite{kacperskiThreegammaAnnihilationImaging2004, kacperskiPerformanceThreephotonPET2005, abuelhiaThreephotonAnnihilationPET2007}.
Initial feasibility was demonstrated through simulations and proof-of-principle measurements with HP-Ge and NaI(Tl) detectors.
More recently, a similar approach was used to produce the first three-to-two ratio image of point sources with GAGG scintillator detectors and $^{18}$F-FDG, achieving a spatial resolution of approximately 1.1\,cm without tomographic reconstruction or \ac{TOF} information~\cite{fujimotoAdvancingPETDirect2025}.
The energy-based approach requires energy resolutions of a few per cent, which are not achieved by current clinical \ac{PET} systems.
The second class employs a time-based trilateration strategy which reconstructs the vertex from photon arrival times and interaction positions alone~\cite{gajosTrilaterationbasedReconstructionOrthopositronium2017}.
While relaxing the energy resolution requirement, this approach currently achieves a spatial resolution of approximately 8\,cm~\cite{moskalTestingCPTSymmetry2021b}.

The remainder of this article is organised as follows.
\Cref{sec:prior} derives the physical constraints imposed by energy-momentum conservation on candidate vertex positions, establishes the triangle condition, and discusses both uninformative and \ac{QED}-informed prior distributions over the decay vertex.
\Cref{sec:reconstructionEnergy} presents the closed-form analytical solution of the energy-based vertex reconstruction algorithm.

\section{Physical constraints on the three-photon decay and prior distribution over the annihilation vertex}
\label{sec:prior}

Let us consider the decay of \ac{o-Ps} at rest into three photons described by the momenta $\vec{\omega}_1, \vec{\omega}_2, \vec{\omega}_3$. 
The energy of the annihilation photons may be written as $E_{i} =\hbar \omega_{i}$ for $i \in \{1,2,3\}$ with $\omega_{i} = |\vec{\omega}_i|$ and without loss of generality $\hbar$ is set to 1.
Therefore, the momentum-energy conservation equations are:

\begin{align}
    \label{eq:momentum_conservation}
\vec{\omega}_1 + \vec{\omega}_2 + \vec{\omega}_3 &= \vec{0} \\
\label{eq:energy_conservation}
\omega_1 + \omega_2 + \omega_3 &= 2 m_e c^2    
\end{align} 
as the total energy of the \ac{o-Ps} at rest is equal to the sum of masses of its constituent $e^{-}$ and $e^{+}$,  equivalent to twice the mass of the electron, where we neglect the binding energy of a few eV. Subsequently, all derivations are given in the \ac{o-Ps} rest frame. 

\Cref{eq:momentum_conservation} implies that any momentum vector lies in the plane spanned by the remaining two, so all three vectors are coplanar. Since photon tracks are straight lines from the decay point $\mathbf{x}$, the decay point itself lies in the same plane. Assuming the hit positions $\vec{P}_i$ for $i \in \{1,2,3\}$ are measured with high precision, they define the decay plane and reduce the reconstruction problem to a 2-D space.

Let us comment on the assumption that we consider \ac{o-Ps} to be at rest, and therefore we know its total energy beforehand.   In practice, this assumption is well justified, e.g. in tissues since the probability of \ac{Ps} formation depends on the positron kinetic energy and it is strongly peaked near smaller values below 50 eV~\cite{championMovingOrganDose2005}.  

\subsection{Triangle condition on the decay vertex}

\begin{figure}[htbp]
\centering
  \begin{subfigure}[t]{0.47\textwidth}
  \centering
\begin{tikzpicture}[
    >=Stealth,
    font=\small,
    hit/.style={circle, fill=hitblue, draw=white, line width=0.6pt,
                inner sep=0pt, minimum size=7pt},
    vertex/.style={circle, inner sep=0pt, minimum size=7pt},
    unitvec/.style={->, line width=1.0pt},
    track/.style={line width=0.8pt, dashed},
]

  \draw[gray!40, line width=1pt, dashed] (0,0) circle (2.8cm);

  \coordinate (H1) at (50:2.8);
  \coordinate (H2) at (170:2.8);
  \coordinate (H3) at (290:2.8);

  \coordinate (V) at (0.25, -0.20);

  \fill[validgreen, opacity=0.12] (H1) -- (H2) -- (H3) -- cycle;

  \draw[validgreen, line width=1.0pt] (H1) -- (H2) -- (H3) -- cycle;

  \draw[validgreen!70, track] (V) -- (H1);
  \draw[validgreen!70, track] (V) -- (H2);
  \draw[validgreen!70, track] (V) -- (H3);

  \draw[validgreen, ->, line width=1.2pt, shorten >=2pt]
      (V) -- ($(V)!0.55!(H1)$) node[right, yshift=2pt] {$\vec{\omega}_1$};
  \draw[validgreen, ->, line width=1.2pt, shorten >=2pt]
      (V) -- ($(V)!0.55!(H2)$) node[left,  yshift=2pt] {$\vec{\omega}_2$};
  \draw[validgreen, ->, line width=1.2pt, shorten >=2pt]
      (V) -- ($(V)!0.55!(H3)$) node[right, yshift=-4pt] {$\vec{\omega}_3$};

  \draw[gray!60, line width=0.6pt]
      (V) ++(50:0.55) arc (50:170:0.55)
      node[midway, above left, gray, font=\footnotesize] {$\alpha_{12}$};
  \draw[gray!60, line width=0.6pt]
      (V) ++(170:0.42) arc (170:290:0.42)
      node[midway, left, gray, font=\footnotesize] {$\alpha_{23}$};
  \draw[gray!60, line width=0.6pt]
      (V) ++(290:0.32) arc (290:410:0.32)
      node[midway, below right, gray, font=\footnotesize] {$\alpha_{31}$};

  \node[hit] at (H1) {};
  \node[above right=2pt, hitblue] at (H1) {$\vec{P}_1$};
  \node[hit] at (H2) {};
  \node[above left=2pt, hitblue] at (H2) {$\vec{P}_2$};
  \node[hit] at (H3) {};
  \node[below=3pt, hitblue] at (H3) {$\vec{P}_3$};

  \node[vertex, fill=validgreen, draw=white, line width=0.6pt] at (V) {};
  \node[above right=2pt, validgreen] at (V) {$\mathbf{x}$};

  \begin{scope}[xshift=2.2cm, yshift=-3.55cm, scale=0.55]
    \draw[gray!35, line width=0.5pt] (0,0) circle (1cm);
    \coordinate (n1a) at (50:1);
    \coordinate (n2a) at (170:1);
    \coordinate (n3a) at (290:1);
    \fill[validgreen, opacity=0.25] (n1a) -- (n2a) -- (n3a) -- cycle;
    \draw[validgreen!60, line width=0.5pt] (n1a) -- (n2a) -- (n3a) -- cycle;
    \draw[hitblue, unitvec, line width=0.8pt] (0,0) -- (n1a)
        node[above right, font=\tiny, hitblue] {$\hat{n}_1$};
    \draw[hitblue, unitvec, line width=0.8pt] (0,0) -- (n2a)
        node[above left,  font=\tiny, hitblue] {$\hat{n}_2$};
    \draw[hitblue, unitvec, line width=0.8pt] (0,0) -- (n3a)
        node[below,       font=\tiny, hitblue] {$\hat{n}_3$};
    \fill[validgreen] (0,0) circle (2pt);
    \node[font=\tiny, validgreen, below right=1pt] at (0,0) {$\mathbf{0}$};
  \end{scope}

\end{tikzpicture}
  \caption{Valid vertex ($\max\,\alpha_{ij} < \pi$): $\mathbf{x} \in \triangle\vec{P}_1\vec{P}_2\vec{P}_3$.
    Unit vectors surround $\mathbf{x}$; their weighted sum can vanish, satisfying momentum conservation with all $\omega_i > 0$.}
  \label{fig:valid_triangle}
  \end{subfigure}
  \hfill
  \begin{subfigure}[t]{0.47\textwidth}
  \centering
\begin{tikzpicture}[
    >=Stealth,
    font=\small,
    hit/.style={circle, fill=hitblue, draw=white, line width=0.6pt,
                inner sep=0pt, minimum size=7pt},
    vertex/.style={circle, inner sep=0pt, minimum size=7pt},
    unitvec/.style={->, line width=1.0pt},
    track/.style={line width=0.8pt, dashed},
]

  \draw[gray!40, line width=1pt, dashed] (0,0) circle (2.8cm);

  \coordinate (H1) at (50:2.8);
  \coordinate (H2) at (170:2.8);
  \coordinate (H3) at (290:2.8);

  \coordinate (V) at (-1.1, 1.6);

  \fill[validgreen, opacity=0.08] (H1) -- (H2) -- (H3) -- cycle;
  \draw[validgreen!40, line width=0.8pt] (H1) -- (H2) -- (H3) -- cycle;

  \draw[invalidred!60, track] (V) -- (H1);
  \draw[invalidred!60, track] (V) -- (H2);
  \draw[invalidred!60, track] (V) -- (H3);

  \draw[invalidred, ->, line width=1.2pt, shorten >=2pt]
      (V) -- ($(V)!0.50!(H1)$) node[right, yshift=2pt]  {$\vec{\omega}_1$};
  \draw[invalidred, ->, line width=1.2pt, shorten >=2pt]
      (V) -- ($(V)!0.50!(H2)$) node[left,  yshift=2pt]  {$\vec{\omega}_2$};
  \draw[invalidred, ->, line width=1.2pt, shorten >=2pt]
      (V) -- ($(V)!0.50!(H3)$) node[right, yshift=-4pt] {$\vec{\omega}_3$};

  \draw[invalidred, line width=1.2pt]
      (V) ++(10.6:0.65) arc (10.6:213.9:0.65) node[invalidred, font=\footnotesize, anchor=south east,midway] {$\alpha_{12} > \pi$};

  \draw[gray!60, line width=0.6pt]
      (V) ++(213.9:0.42) arc (213.9:295.9:0.42)
      node[midway, below left, gray, font=\footnotesize] {$\alpha_{23}$};
  \draw[gray!60, line width=0.6pt]
      (V) ++(295.9:0.32) arc (295.9:370.6:0.32)
      node[midway, below right, gray, font=\footnotesize] {$\alpha_{31}$};

  \draw[invalidred, line width=1.0pt, dash pattern=on 4pt off 2pt]
      (-2.625, 0.975) -- (1.185, 2.537);

  \node[hit] at (H1) {};
  \node[above right=2pt, hitblue] at (H1) {$\vec{P}_1$};
  \node[hit] at (H2) {};
  \node[above left=2pt, hitblue] at (H2) {$\vec{P}_2$};
  \node[hit] at (H3) {};
  \node[below=3pt, hitblue] at (H3) {$\vec{P}_3$};

  \node[vertex, fill=invalidred, draw=white, line width=0.6pt] at (V) {};
  \node[above right=2pt, invalidred] at (V) {$\mathbf{x}$};

  \begin{scope}[xshift=2.2cm, yshift=-3.55cm, scale=0.55]
    \draw[gray!35, line width=0.5pt] (0,0) circle (1cm);
    \coordinate (n1b) at (8:1);
    \coordinate (n2b) at (211:1);
    \coordinate (n3b) at (289:1);
    \fill[invalidred, opacity=0.15] (n1b) -- (n2b) -- (n3b) -- cycle;
    \draw[invalidred!60, line width=0.5pt] (n1b) -- (n2b) -- (n3b) -- cycle;
    \draw[hitblue, unitvec, line width=0.8pt] (0,0) -- (n1b)
        node[right,      font=\tiny, hitblue] {$\hat{n}_1$};
    \draw[hitblue, unitvec, line width=0.8pt] (0,0) -- (n2b)
        node[below left, font=\tiny, hitblue] {$\hat{n}_2$};
    \draw[hitblue, unitvec, line width=0.8pt] (0,0) -- (n3b)
        node[below,      font=\tiny, hitblue] {$\hat{n}_3$};
    \fill[invalidred] (0,0) circle (2pt);
    \node[font=\tiny, invalidred, above right=1pt] at (0,0) {$\mathbf{0}$};
  \end{scope}
\end{tikzpicture}
  \caption{Invalid vertex ($\max\,\alpha_{ij} > \pi$): $\mathbf{x} \notin \triangle\vec{P}_1\vec{P}_2\vec{P}_3$.
    All unit vectors lie within an open half-plane (separated by the dashed line); no positive linear combination can vanish.}
  \label{fig:invalid_triangle}
  \end{subfigure}
  \caption{Triangle condition for candidate vertex $\mathbf{x}$: a physical solution to momentum conservation requires $\mathbf{x}$ to lie strictly inside $\triangle\vec{P}_1\vec{P}_2\vec{P}_3$.}
  \label{fig:triangle_cases}
\end{figure}
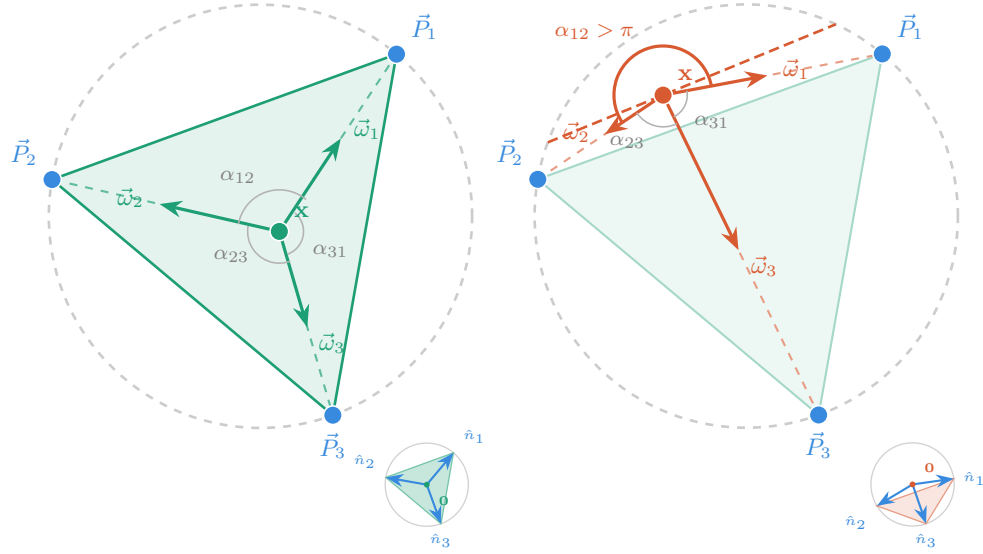

Momentum conservation imposes a further geometric constraint on the candidate vertex $\mathbf{x}$ within the decay plane.
Defining the unit vectors from $\mathbf{x}$ toward each hit as
\begin{equation}
    \vec{n}_i(\mathbf{x}) = \frac{\vec{P}_i - \mathbf{x}}{|\vec{P}_i - \mathbf{x}|},
\end{equation}
and writing $\vec{\omega}_i = \omega_i \vec{n}_i(\mathbf{x})$ for massless photons, \cref{eq:momentum_conservation} becomes:
\begin{equation}
    \label{eq:momentum_unit}
    \sum_{i=1}^{3} \omega_i \vec{n}_i(\mathbf{x}) = \vec{0}, 
    \quad \omega_i > 0.
\end{equation}
A solution with all $\omega_i > 0$ exists if and only if the directions $\vec{n}_i(\mathbf{x})$ surround $\mathbf{x}$, i.e., no straight line through $\mathbf{x}$ places all three hits on the same side.
Equivalently, the largest opening angle between consecutive directions as seen from $\mathbf{x}$ must satisfy:
\begin{equation}
    \max_{i \neq j}\, \alpha_{ij}(\mathbf{x}) < \pi.
    \label{eq:triangle_condition}
\end{equation}
The set of all $\mathbf{x}$ satisfying~\eqref{eq:triangle_condition} is exactly the interior of the triangle $\triangle\vec{P}_1\vec{P}_2\vec{P}_3$.
Outside, all directions $\vec{n}_i$ lie within an open half-plane and their positive linear combination cannot vanish (see \cref{fig:valid_triangle,fig:invalid_triangle}).

\subsection{Role of energy conservation and detector measurements}

Energy conservation~\eqref{eq:energy_conservation} adds one scalar equation to the geometric constraints.
Its practical effect depends on which quantities are measured, and it is instructive to consider two cases.

\textbf{(i) No detector energy measurements.} For each candidate $\mathbf{x}$ inside the triangle, momentum conservation already fixes the ratios $\omega_1 : \omega_2 : \omega_3$ from geometry consideration alone.
Since the \ac{Ps} mass$M =2m_e c^2$ is known a priori,  it fixes the overall energy scale, making a definite prediction $\tilde{\omega}_i(\mathbf{x})$ for each photon energy at each $\mathbf{x}$.
Every interior point remains geometrically reachable,  but it converts each candidate vertex into a testable energy prediction.

\textbf{(ii) Energy measurements available.} When the detector measures photon energies $E_i$, these must be consistent with the predicted values $\tilde{\omega}_i(\mathbf{x})$.
With perfect measurements, this selects a unique point $\mathbf{x}^*$ inside the triangle.
Under finite detector energy resolution $\sigma_E$, the constraint broadens into a likelihood.

Let us note that the energy measurement constrains the directions $\vec{n}_i(\mathbf{x})$ and thereby the ratios of photon momenta. In contrast, the \ac{TOF} measurements constrain the distances $|\vec{P}_i - \mathbf{x}|$, placing $\mathbf{x}$ on circles centred on each detector hit. The two measurement types are therefore geometrically complementary.

\subsection{Prior distribution}

Since the hit positions $\vec{P}_i$ are treated as precisely known for our reconstruction, all consequences of energy-momentum conservation are fully determined before any energy or timing measurement is made.
They therefore constitute prior knowledge about the candidate vertex $\mathbf{x}$, independent of the detector responses.
Coplanarity, the triangle condition, and the QED-based decay dynamics can, therefore, be treated as a part of the prior.

The first constraint is of a geometrical nature: a physical solution requires $\mathbf{x}$ to lie inside $\triangle\vec{P}_1\vec{P}_2\vec{P}_3$, as established in~\eqref{eq:triangle_condition}.
This defines the support of the prior.

The second constraint comes from the QED decay dynamics: given a candidate $\mathbf{x}$, the predicted energies $\tilde{\omega}_i(\mathbf{x})$ are fixed by geometry via the sine rule (see \cref{sec:reconstructionEnergy}), and their probability is given by the Ore--Powell matrix element~\cite{oreThreePhotonAnnihilationElectronPositron1949,beresteckijQuantumElectrodynamics2008}:
\begin{equation}
    P(\tilde{\omega}_1, \tilde{\omega}_2) \propto \sum_{i=1}^3
    \left(\frac{m_e c^2 - \tilde{\omega}_i}{\tilde{\omega}_j\, \tilde{\omega}_k}\right)^2,
    \quad \{i,j,k\} = \{1,2,3\}
    \label{eq:ore_powell}
\end{equation}

\begin{figure}[htb]
\centerline {
\includegraphics[width=10cm]{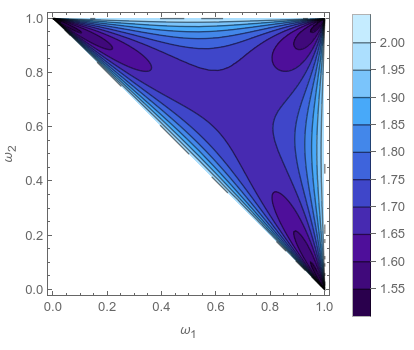}}
\caption{The unnormalized probability distribution $P({\tilde{\omega}}_1, \tilde{{\omega}}_2)$ over the kinematically accessible phase space, shown as a function of normalised photon energies $\tilde{\omega}_i/m_{e}c^2 \in (0,1)$.}
\label{fig:dalitz}
\end{figure}

Equation~\eqref{eq:ore_powell}, illustrated in \cref{fig:dalitz}, modulates the prior within its support:
\begin{equation}
    P(\mathbf{x}) = \kappa \sum_{i=1}^{3}
    \left(\frac{m_e c^2 - \tilde{\omega}_i(\mathbf{x})}
    {\tilde{\omega}_j(\mathbf{x})\,\tilde{\omega}_k(\mathbf{x})}\right)^2
    \cdot\, \mathbf{1}_{\mathbf{x}\,\in\,\triangle\vec{P}_1\vec{P}_2\vec{P}_3},
    \quad i,j,k \in \{1,2,3\},
    \label{eq:full_prior}
\end{equation}
where $\kappa$ is a normalisation constant.
The indicator function $\mathbf{1}_{\mathbf{x}\,\in\,\triangle}$ enforces the hard geometric boundary, while the Ore-Powell factor assigns a higher probability to vertex positions corresponding to kinematically favoured energy configurations, independently of the measured times and energies.
Replacing the Ore-Powell factor by a constant recovers the flat Dalitz baseline~\cite{dalitzCXIIAnalysisTmeson1953}, representing genuine ignorance about the energy sharing among the three photons.

\section{Energy-based position reconstruction derivation}
\label{sec:reconstructionEnergy}

\begin{figure}[htb]
\centerline {
\includegraphics[width=10cm]{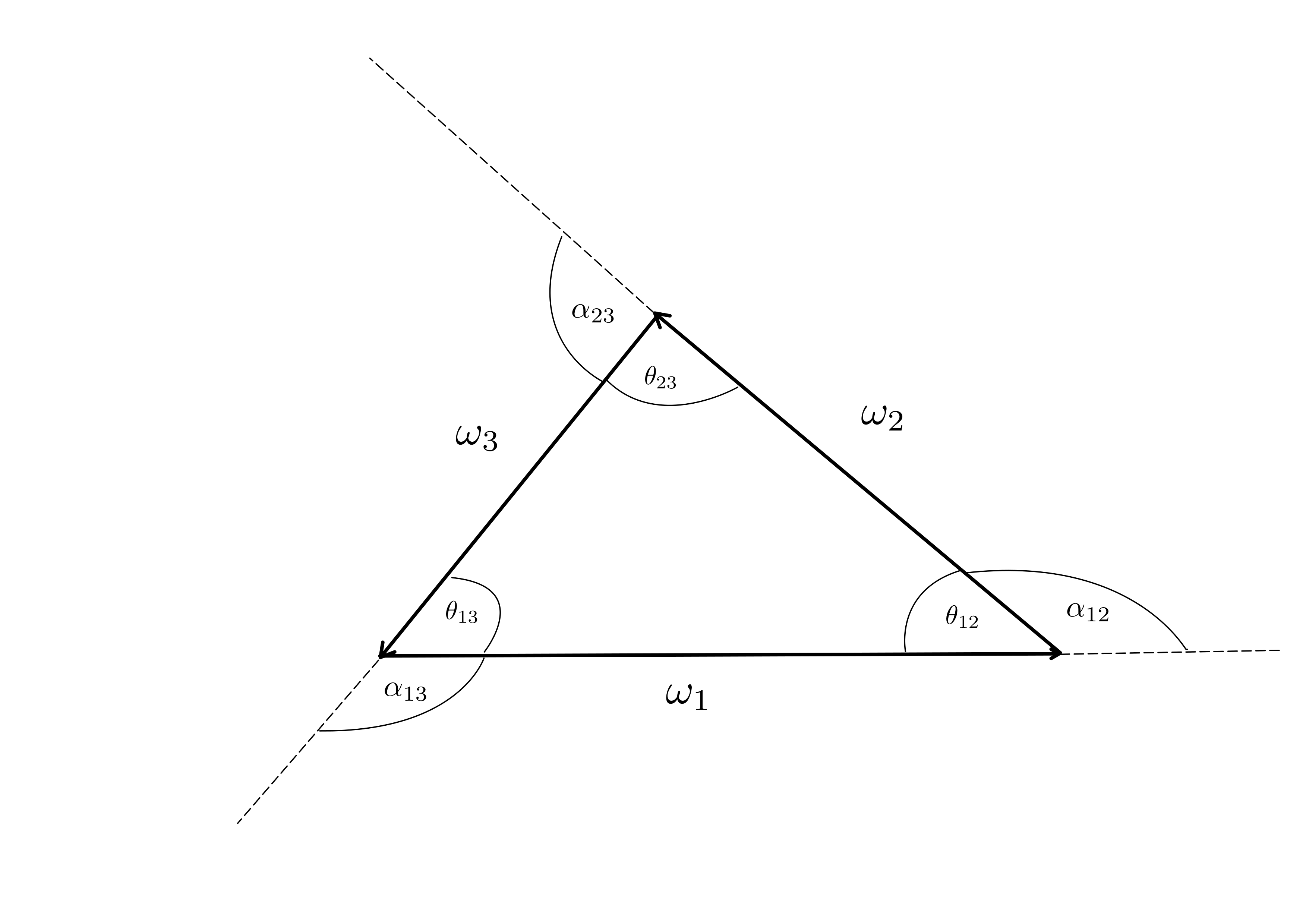}}
\caption{Relation between angles in the triangle defined by the momentum
vectors ($\theta_{ij}$) and the opening angles ($\alpha_{ij}$).}
\label{fig:triangle_rel}
\end{figure}

In the following, we present a derivation of the exact solution of the energy-based \ac{o-Ps} position reconstruction.
The vector sum of $\omega_i$ elements expresses the momentum conservation law (see Eq.~\ref{eq:momentum_conservation}), and may be represented by a triangle with angles $\theta_{12}+\theta_{13}+ \theta_{23} = \pi$ calculated using the cosine rule:
\begin{equation}\label{eq:theta_cosine_rule}
	\left(2 m_e c^2 - \omega_i  - \omega_j \right)^2 = \omega_i^2 + \omega_j^2 - 2\,\omega_i\omega_j\cos\theta_{ij}, \quad 1 \leq i < j \leq 3.
\end{equation}
The dependence between the angles in the triangle defined by the momentum vectors ($\theta_{ij}$) and the opening angles ($\alpha_{ij}$) is given by:
\begin{equation}
	\theta_{ij} +\alpha_{ij} =  \pi	\quad\quad 1 \leq i < j \leq 3.    \label{eq:theta_alpha}
\end{equation}
and it is illustrated in Fig.~\ref{fig:triangle_rel}.

\begin{figure}
	\centering
	\begin{tikzpicture}[scale=.95]

		\def\clipSize{4.5}
		\clip (-\clipSize,-\clipSize) rectangle (\clipSize,\clipSize) ;

		\draw[axis] (-4,3) -- (4,3) node [right] {$x''$};
		\draw[axis] (0,-4) -- (0,4) node [above] {$y''$};

		\coordinate (A) at (0, 0);
		\coordinate (H1) at (0, 3.0);
		\coordinate (H2) at (3.2, -3.2);
		\coordinate (H3) at (-2.7,-2.3);
		\coordinate (OY) at (0, 4);

		\draw[black, thick] (H1)  -- (H2) node[right,midway] {$d_{12}$};
		\draw[black, thick] (H2) --  (H3) node[below,midway] {$d_{23}$};
		\draw[black, thick] (H1) --  (H3) node[left,midway] {$d_{13}$};

		\draw[densely dotted, thick] (A) -- (-4,4) node [left, midway] {$y'' = a_2 x'' -r_1$};
		\draw[densely dotted, thick] (A) -- (4.05,3.45) node [right, midway] {$y'' = a_3 x'' -r_1$};
		\draw[positron] (A) -- (H1) node[right,midway] {$r_{1}$} node [above left] {$P_1(0,0)$};
		\draw[positron] (A) -- (H2) node[above,midway] {$r_{2}$} node [below] {$P_2(x_2'',y_2'')$};
		\draw[positron] (A) -- (H3) node[above,midway] {$r_{3}$} node [below] {$P_3(x_3'',y_3'')$};

		\def\angleRadius{0.75}

		\draw[blue] let \p1=(H1), \p2=(OY), \p3=(H2),
		\n1={atan2(\y2-\y1,\x2-\x1)}, \n2={atan2(\y3-\y1,\x3-\x1)} in
		($(H1)+(\n1:\angleRadius)$) arc (\n1:\n2:\angleRadius);
		\draw[blue] let \p1=(H1), \p2=(OY), \p3=(H2),
		\n1={atan2(\y2-\y1,\x2-\x1)}, \n2={atan2(\y3-\y1,\x3-\x1)} in
		($(H1)+(0.5*\n1+0.5*\n2:\angleRadius*0.5)$) node[anchor=center]{$\gamma$};

		\draw[blue] let \p1=(H1), \p2=(H2), \p3=(H3),
		\n1={atan2(\y2-\y1,\x2-\x1)}, \n2={atan2(\y3-\y1,\x3-\x1)} in
		($(H1)+(\n1:\angleRadius*1.2)$) arc (\n1:\n2:\angleRadius*1.2);
		\draw[blue] let \p1=(H1), \p2=(H2), \p3=(H3),
		\n1={atan2(\y2-\y1,\x2-\x1)}, \n2={atan2(\y3-\y1,\x3-\x1)} in
		($(H1)+(0.5*\n1+0.5*\n2:\angleRadius*0.75)$) node[anchor=center]{$\gamma_{23}$};

		\draw[blue] let \p1=(H2), \p2=(H1), \p3=(H3),
		\n1={atan2(\y2-\y1,\x2-\x1)}, \n2={atan2(\y3-\y1,\x3-\x1)} in
		($(H2)+(\n1:\angleRadius*1.2)$) arc (\n1:\n2:\angleRadius*1.2);
		\draw[blue] let \p1=(H2), \p2=(H1), \p3=(H3),
		\n1={atan2(\y2-\y1,\x2-\x1)}, \n2={atan2(\y3-\y1,\x3-\x1)} in
		($(H2)+(0.5*\n1+0.5*\n2:\angleRadius*0.75)$) node[anchor=center]{$\gamma_{13}$};

		\draw[blue] let \p1=(H3), \p2=(H1), \p3=(H2),
		\n1={atan2(\y2-\y1,\x2-\x1)}, \n2={atan2(\y3-\y1,\x3-\x1)} in
		($(H3)+(\n1:\angleRadius*1.2)$) arc (\n1:\n2:\angleRadius*1.2);
		\draw[blue] let \p1=(H3), \p2=(H1), \p3=(H2),
		\n1={atan2(\y2-\y1,\x2-\x1)}, \n2={atan2(\y3-\y1,\x3-\x1)} in
		($(H3)+(0.5*\n1+0.5*\n2:\angleRadius*0.75)$) node[anchor=center]{$\gamma_{12}$};

		\draw[blue] let \p1=(A), \p2=(H1), \p3=(H2),
		\n1={atan2(\y2-\y1,\x2-\x1)}, \n2={atan2(\y3-\y1,\x3-\x1)} in
		($(A)+(\n1:\angleRadius)$) arc (\n1:\n2:\angleRadius);
		\draw[blue] let \p1=(A), \p2=(H1), \p3=(H2),
		\n1={atan2(\y2-\y1,\x2-\x1)}, \n2={atan2(\y3-\y1,\x3-\x1)} in
		($(A)+(0.5*\n1+0.5*\n2:\angleRadius*0.5)$) node[anchor=center]{$\alpha_{12}$};

		\draw[blue] let \p1=(A), \p2=(H2), \p3=(H3),
		\n1={atan2(\y2-\y1,\x2-\x1)}, \n2={atan2(\y3-\y1,\x3-\x1)} in
		($(A)+(\n1:\angleRadius)$) arc (\n1:\n2:\angleRadius);
		\draw[blue] let \p1=(A), \p2=(H2), \p3=(H3),
		\n1={atan2(\y2-\y1,\x2-\x1)}, \n2={atan2(\y3-\y1,\x3-\x1)} in
		($(A)+(0.5*\n1+0.5*\n2:\angleRadius*0.65)$) node[anchor=center]{$\alpha_{23}$};

		\draw[blue] let \p1=(A), \p2=(H3), \p3=(H1),
		\n1={atan2(\y2-\y1,\x2-\x1)}, \n2={atan2(\y3-\y1,\x3-\x1)} in
		($(A)+(\n1:\angleRadius)$) arc (\n1:\n2-360:\angleRadius);
		\draw[blue] let \p1=(A), \p2=(H1), \p3=(H3),
		\n1={atan2(\y2-\y1,\x2-\x1)}, \n2={atan2(\y3-\y1,\x3-\x1)} in
		($(A)+(0.5*\n1+0.5*\n2-180:\angleRadius*0.5)$) node[anchor=center]{$\alpha_{13}$};

	\end{tikzpicture}
	\caption{Reconstruction of decay position using energy (equivalently angle) constraints.}
	\label{fig:triangle_xb}
\end{figure}
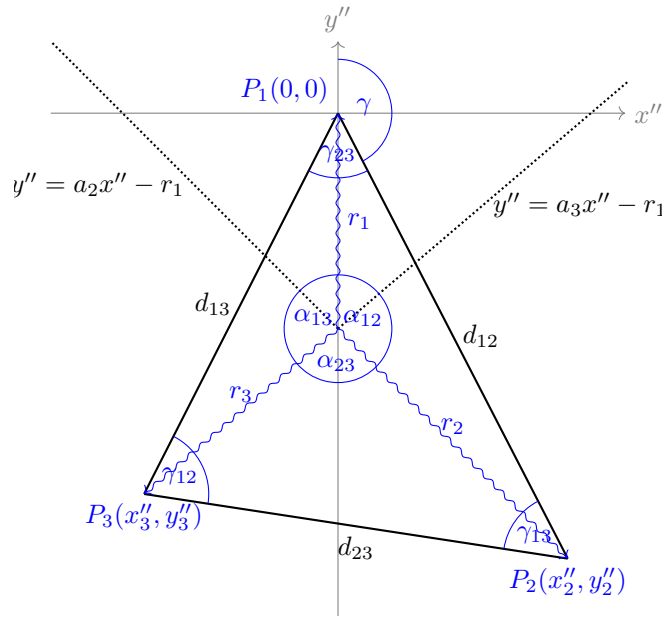

We analyze data in a shifted and rotated decay plane ($x'',y''$).
The points $P_1$, $P_2$ and $P_3$ corresponding to the coordinates of the three registered hits are situated on the decay plane ($x'',y''$) as shown in \cref{fig:triangle_xb}.
The origin of the ($x'',y''$) plane is attached to the $P_1$ point that is the vertex of the triangle $P_1,P_2,P_3$ lying at the smallest angle $\gamma_{23}$:
\begin{equation}
	\label{eq:gamma_23}
	\gamma_{23} = \min \left(\gamma_{12}, \gamma_{13}\right).
\end{equation}
Under the assumption that the decay point is on the $y''$-axis with a negative ordinate value (see \cref{fig:triangle_xb}), the rotation angle of the triangle ($\gamma$) is in the second quadrant.
Taking into account additionally \cref{eq:gamma_23}, the limits of $\gamma$ are:
\begin{equation*}
	\frac{2\pi}{3} \leq \gamma \leq \pi.
\end{equation*}
The rotation angle of the triangle ($\gamma$) as well as the distances between the points $P_i$ and decay position ($r_i$) for $i \in \{1, 2, 3\}$ are unknown.
This derivation will be focused only on the estimation of the $r_1$ distance that is used to parameterize the lines
\begin{align}
	y'' & = a_2 \, x'' - r_1   \label{eq:lines_a2} \\
	y'' & = a_3 \, x'' - r_1  \label{eq:lines_a3}
\end{align}
passing through the decay position and points $P_2$ and $P_3,$ respectively.
The slopes of the lines $a_2$ and $a_3$ can be calculated based on opening angles $\alpha_{12}$ and $\alpha_{13},$ respectively as:
\begin{align}
	a_2 & = \tan\left(\frac{\pi}{2} -  \alpha_{12} \right)  \\
	a_3 & = \tan\left(\frac{\pi}{2} +  \alpha_{13} \right),
\end{align}
where the periodicity of $\pi$ of the tangent function was taken into account.
It is evident that the points $P_2$ and $P_3$ in polar representation depend only on the single unknown $\gamma$, describing the rotation of the triangle, i.e.
\begin{align}
	\label{eq:polar_p2_p3}
	\begin{split}
		x''_2 = d_{12} \sin \gamma  \quad\quad&x''_3 = d_{13} \sin \left(\gamma + \gamma_{23}\right) \\
		y''_2 = d_{12} \cos \gamma  \quad\quad&y''_3 = d_{13} \cos \left(\gamma + \gamma_{23}\right).
	\end{split}
\end{align}
Subtracting the line equation in \cref{eq:lines_a3} for ($x''_3,y''_3$) from the line equation in \cref{eq:lines_a2} for ($x''_2,y''_2$) defines the linear dependence of the cartesian coordinates ($x''_2,y''_2$) and ($x''_3,y''_3$):
\begin{equation}
	a_2 \,x''_2 - y''_2 = a_3 \,x''_3 - y''_3.	\label{eq:linear_dependence}
\end{equation}
After substitution of polar coordinates from \cref{eq:polar_p2_p3,eq:linear_dependence} it is possible to evaluate the rotation angle $\gamma$:
\begin{equation}
	\label{eq:tan_gamma}
	\gamma =  \pi + \arctan \left(\frac{d_{12} + d_{13} \left(a_3 \sin\gamma_{23} - \cos\gamma_{23} \right)}{a_2 \,d_{12} - d_{13} \left(a_3 \cos\gamma_{23} + \sin\gamma_{23} \right)}\right).
\end{equation}
The additional shift of $\pi$ is due to the value of the $\arctan$ function in the range from $-\frac{\pi}{2}$ to $\frac{\pi}{2};$ as mentioned $\gamma$ is in the second quadrant (see \cref{fig:triangle_xb}).
Next, the distances $r_1$, $r_2$ and $r_3$ may be evaluated using sine rule:
\begin{align*}
	r_1 & = d_{12}\frac{\sin\left(\gamma - \alpha_{12}\right)}{\sin\alpha_{12}}        \\
	r_2 & = d_{12}\frac{\sin\left(\pi-\gamma\right)}{\sin\alpha_{12}}                  \\
	r_3 & = d_{13}\frac{\sin\left(\gamma + \gamma_{23} - \pi\right)}{\sin\alpha_{13}}.
\end{align*}

With the distances $r_1, r_2, r_3$ at hand, the \ac{o-Ps} position ($\mathbf{x_E}$) may be found in the original decay plane ($x,y$), as a solution of the non-linear system of equations:
\begin{equation}
	(x_E - x_i)^2 + (y_E - y_i)^2 = r_i^2,\quad i \in \{1, 2, 3\}
	\label{eq:three_circles}
\end{equation}
describing three circles $O_i$ each with radii $r_i$ for $i \in \{1, 2, 3\}$.
The unique solution ($x_E, y_E$) may be obtained directly by solving the linear system of equations
\begin{equation}
	\mathbf{H} \begin{pmatrix}
		x_E \\
		y_E
	\end{pmatrix} = \mathbf{r}
	\label{eq:linear_system}
\end{equation}
where
\begin{equation}
	\mathbf{H} = 2\begin{pmatrix}
		x_1 - x_2 & y_1 - y_2 \\
		x_1 - x_3 & y_1 - y_3
	\end{pmatrix}
	\label{eq:matrix_h}
\end{equation}
and
\begin{equation}
	\mathbf{r} = \begin{pmatrix}
		r_2^2 - r_1^2 + x_1^2 - x_2^2 + y_1^2 - y_2^2 \\
		r_3^2 - r_1^2 + x_1^2 - x_3^2 + y_1^2 - y_3^2
	\end{pmatrix}
	\label{eq:vector_r}
\end{equation}
evaluated based on \cref{eq:three_circles} by subtracting the circle equation $O_j$ from circle equation $O_i$ for $i < j.$

\section{Conclusion}

We have analysed the physical constraints imposed by energy-momentum conservation and the QED-based matrix element on the three-photon decay of ortho-positronium and discussed their implications for annihilation vertex reconstruction.

Momentum conservation alone establishes two nested geometrical constraints:
coplanarity reduces the reconstruction problem from three to two dimensions, and
the triangle condition confines all physically admissible candidate vertices to the interior of the triangle formed by the three photon hit positions.
These constraints are determined entirely by the measured hit positions and are independent of any energy or timing measurement.

Energy conservation, combined with the known positronium mass, converts each candidate vertex inside the triangle into a unique, testable prediction for the three photon energies via the sine rule (see ~\cref{sec:reconstructionEnergy}).
When detector energy measurements are available, this prediction selects a unique true vertex under perfect resolution.
\Ac{TOF} measurements provide geometrically complementary information, constraining distances rather than directions.

The QED decay dynamics, described by the Ore-Powell matrix element, modulate the prior within the triangular support.
The resulting distribution, shown in \cref{fig:dalitz}, enhances the probability of configurations in which one photon is soft (carries little energy), corresponding to the edges of the Dalitz plot where $\omega_i \to 0$.

Finally, we have presented a closed-form analytical derivation of the energy-based vertex reconstruction algorithm.
The algorithm exploits the angular constraints imposed by momentum conservation to reduce the problem to a linear system, yielding a unique solution for the annihilation vertex in the decay plane without requiring iterative optimisation.
\printbibliography

\end{document}